\documentclass{fac}

\usepackage{hyperref}
\usepackage{graphicx}
\usepackage{amssymb}
\usepackage{amsfonts}
\usepackage{amsmath}
\usepackage{dsfont}
\usepackage{MnSymbol}
\usepackage{mathtools}
\usepackage{epsfig}
\usepackage{epstopdf}
\usepackage{lmerge}
\usepackage{tikz}
\usepackage{amsthm}
\ifprodtf \else \usepackage{latexsym}\fi

\usetikzlibrary{calc,decorations.pathmorphing,shapes,graphs}

\newcounter{sarrow}

\newcounter{sarrow1}
\newcommand\xnrsquigarrow[1]{%
\stepcounter{sarrow1}%
\mathrel{\begin{tikzpicture}[baseline= {( $ (current bounding box.south) + (0,-0.5ex) $ )}]
\node[inner sep=.5ex] (\thesarrow) {$\scriptstyle #1$};
\path[draw,<-,decorate,
  decoration={zigzag,amplitude=0.7pt,segment length=1.2mm,pre=lineto,pre length=4pt}]
    (\thesarrow1.south east) -- (\thesarrow1.south west);
    $\slashedarrowfill@\relbar\relbar/$
    \end{tikzpicture}}%
}

\makeatletter
\def\slashedarrowfill@#1#2#3#4#5{%
  $\m@th\thickmuskip0mu\medmuskip\thickmuskip\thinmuskip\thickmuskip
   \relax#5#1\mkern-7mu%
   \cleaders\hbox{$#5\mkern-2mu#2\mkern-2mu$}\hfill
   \mathclap{#3}\mathclap{#2}%
   \cleaders\hbox{$#5\mkern-2mu#2\mkern-2mu$}\hfill
   \mkern-7mu#4$%
}
\def\rightslashedarrowfillb@{%
  \slashedarrowfill@\relbar\relbar/\rightarrow}
\newcommand\xnrightarrow[2][]{%
  \ext@arrow 0055{\rightslashedarrowfillb@}{#1}{#2}}

\def\rightslashedarrowfille@{%
  \slashedarrowfill@\relbar\relbar/\twoheadrightarrow}
\newcommand\xntworightarrow[2][]{%
  \ext@arrow 0055{\rightslashedarrowfille@}{#1}{#2}}

\def\rightslashedarrowfillg@{%
  \slashedarrowfill@\relbar\relbar{\raisebox{.12em}{}}\twoheadrightarrow}
\newcommand\xtworightarrow[2][]{%
  \ext@arrow 0055{\rightslashedarrowfillg@}{#1}{#2}}

\def\rightslashedarrowfillx@{%
  \slashedarrowfill@\Relbar\Relbar/\rightrightarrows}
\newcommand\xnTworightarrow[2][]{%
  \ext@arrow 0055{\rightslashedarrowfillx@}{#1}{#2}}

\def\rightslashedarrowfilly@{%
  \slashedarrowfill@\Relbar\Relbar{\raisebox{.12em}{}}\rightrightarrows}
\newcommand\xTworightarrow[2][]{%
  \ext@arrow 0055{\rightslashedarrowfilly@}{#1}{#2}}

\pgfdeclareshape{slash underlined}
{
  \inheritsavedanchors[from=rectangle] 
  \inheritanchorborder[from=rectangle]
  \inheritanchor[from=rectangle]{north}
  \inheritanchor[from=rectangle]{north west}
  \inheritanchor[from=rectangle]{north east}
  \inheritanchor[from=rectangle]{center}
  \inheritanchor[from=rectangle]{west}
  \inheritanchor[from=rectangle]{east}
  \inheritanchor[from=rectangle]{mid}
  \inheritanchor[from=rectangle]{mid west}
  \inheritanchor[from=rectangle]{mid east}
  \inheritanchor[from=rectangle]{base}
  \inheritanchor[from=rectangle]{base west}
  \inheritanchor[from=rectangle]{base east}
  \inheritanchor[from=rectangle]{south}
  \inheritanchor[from=rectangle]{south west}
  \inheritanchor[from=rectangle]{south east}
  \inheritanchorborder[from=rectangle]
  \foregroundpath{
    \southwest \pgf@xa=\pgf@x \pgf@ya=\pgf@y
    \northeast \pgf@xb=\pgf@x \pgf@yb=\pgf@y
    \pgf@xc=\pgf@xa
    \advance\pgf@xc by .5\pgf@xb
    \pgf@yc=\pgf@ya
    \advance\pgf@xc by -1.3pt
    \advance\pgf@yc by -1.8pt
    \pgfpathmoveto{\pgfqpoint{\pgf@xc}{\pgf@yc}}
    \advance\pgf@xc by  2.6pt
    \advance\pgf@yc by  3.6pt
    \pgfpathlineto{\pgfqpoint{\pgf@xc}{\pgf@yc}}
    \pgfpathmoveto{\pgfqpoint{\pgf@xa}{\pgf@ya}}
    \pgfpathlineto{\pgfqpoint{\pgf@xb}{\pgf@ya}}
 }
}
\tikzset{nomorepostaction/.code=\let\tikz@postactions\pgfutil@empty}

\newtheorem{theorem}{Theorem}[section]

\newtheorem{proposition}[theorem]{Proposition}

\title[Draft of On the Parallel Composition for True Concurrency]
      {On the Parallel Composition for True Concurrency}

\author[Yong Wang]
    {Yong Wang\\
     College of Computer Science and Technology,\\
     Faculty of Information Technology,\\
     Beijing University of Technology, Beijing, China\\
     }

\correspond{Yong Wang, Pingleyuan 100, Chaoyang District, Beijing, China.
            e-mail: wangy@bjut.edu.cn}

\pubyear{2017}
\pagerange{\pageref{firstpage}--\pageref{lastpage}}

\begin{document}
\label{firstpage}

\makecorrespond

\maketitle

\begin{abstract}
For insight into the parallel composition for true concurrency, we recall the axiomatization of the parallel composition modulo truly concurrent behavioral equivalences as the sidelights
of truly concurrent process algebra APTC. We prove that: (1) There is a finite sound and complete axiomatization of the parallel composition modulo pomset, step and hp-bisimulations, 
without any auxiliary operators. (2) There does not exist a finite sound and complete axiomatization of the parallel composition modulo hhp-bisimulation, without any auxiliary operator.
(3) There is a finite sound and complete axiomatization of the parallel composition modulo pomset, step, hp-. and hhp-bisimulations, with the auxiliary left parallel composition
  and communication merge.
\end{abstract}


\section{Introduction}{\label{int}}

True concurrency is characterized by the truly concurrent behavioral equivalences, including pomset bisimulation $\sim_p$, step bisimulation $\sim_s$, history-preserving (hp-) bisimulation $\sim_{hp}$, and hereditary
history-preserving (hhp-) bisimulation $\sim_{hhp}$ (for the details, please refer to \cite{CTC} and \cite{APTC}).

We designed calculus and axiomatization to capture the characteristics for true concurrency, which are called CTC (Calculus for True Concurrency) \cite{CTC} and APTC (Algebra for Processes
in True Concurrency) \cite{APTC}. In APTC, we design a sound and complete axiomatization for the parallel composition $\parallel$, and here we recall it as some sidelights.

For insight into the parallel composition for true concurrency, we recall a subset of APTC including the atomic events, the alternative composition $+$, the sequential composition $\cdot$, 
and the parallel composition $\parallel$, which we call the result algebra PA (Parallel Algebra).

In the following sections, we discuss PA for truly concurrent bisimulations, in section \ref{as}, we discuss PA for step bisimulation, pomset bisimulation, and history-preserving 
bisimulation, which is called PA1; and for hereditary history-preserving bisimulation in section \ref{oahhp} and \ref{ahhp}, which is called PA2.

\section{Axiomatization of Parallel Composition for Pomset, Step, History-Preserving Bisimulations}{\label{as}}

Let $e_1, e_2, e_1', e_2'\in \mathbb{E}$ where $\mathbb{E}$ is the set of atomic events, and let variables $x,y,z$ range over the set of terms for true concurrency, $p,q,s$ range over the set of closed terms. And
$\gamma(e_1,e_2)|e_1,e_2\in \mathbb{E}$ is the communication function. The PA in this section is called PA1.

We give the transition rules of PA1 in Table \ref{TRForPA1}, it is suitable for truly concurrent behavioral equivalences, including pomset bisimulation, step bisimulation,
hp-bisimulation. The operational semantics of parallel composition implies the parallelism and communication and can be unified by the following equation.

$$e_1\parallel e_2\triangleq
\left\{
    \begin{aligned}
        \gamma(e_1,e_2), & \textrm{ if there is a communication between } e_1,e_2\\
        \{e_1,e_2\}, & \textrm{ else.}
    \end{aligned}
\right. 
$$

\begin{center}
    \begin{table}
        $$\frac{}{e\xrightarrow{e}\surd}$$
        $$\frac{x\xrightarrow{e}\surd}{x+ y\xrightarrow{e}\surd} \quad\frac{x\xrightarrow{e}x'}{x+ y\xrightarrow{e}x'} \quad\frac{y\xrightarrow{e}\surd}{x+ y\xrightarrow{e}\surd}
        \quad\frac{y\xrightarrow{e}y'}{x+ y\xrightarrow{e}y'}$$
        $$\frac{x\xrightarrow{e}\surd}{x\cdot y\xrightarrow{e} y} \quad\frac{x\xrightarrow{e}x'}{x\cdot y\xrightarrow{e}x'\cdot y}$$
        $$\frac{x\xrightarrow{e_1}\surd\quad y\xrightarrow{e_2}\surd}{x\parallel y\xrightarrow{\{e_1,e_2\}}\surd} \quad\frac{x\xrightarrow{e_1}x'\quad y\xrightarrow{e_2}\surd}{x\parallel y\xrightarrow{\{e_1,e_2\}}x'}$$
        $$\frac{x\xrightarrow{e_1}\surd\quad y\xrightarrow{e_2}y'}{x\parallel y\xrightarrow{\{e_1,e_2\}}y'} \quad\frac{x\xrightarrow{e_1}x'\quad y\xrightarrow{e_2}y'}{x\parallel y\xrightarrow{\{e_1,e_2\}}x'\parallel y'}$$
        $$\frac{x\xrightarrow{e_1}\surd\quad y\xrightarrow{e_2}\surd}{x\parallel y\xrightarrow{\gamma(e_1,e_2)}\surd} \quad\frac{x\xrightarrow{e_1}x'\quad y\xrightarrow{e_2}\surd}{x\parallel y\xrightarrow{\gamma(e_1,e_2)}x'}$$
        $$\frac{x\xrightarrow{e_1}\surd\quad y\xrightarrow{e_2}y'}{x\parallel y\xrightarrow{\gamma(e_1,e_2)}y'} \quad\frac{x\xrightarrow{e_1}x'\quad y\xrightarrow{e_2}y'}{x\parallel y\xrightarrow{\gamma(e_1,e_2)}x'\parallel y'}$$
        \caption{Transition rules of PA1}
        \label{TRForPA1}
    \end{table}
\end{center}

We can design the axioms of PA1 as Table \ref{AxiomsForPA1} shows.

\begin{center}
    \begin{table}
        \begin{tabular}{@{}ll@{}}
            \hline No. &Axiom\\
            $A1$ & $x+ y = y+ x$\\
            $A2$ & $(x+ y)+ z = x+ (y+ z)$\\
            $A3$ & $x+ x = x$\\
            $A4$ & $(x+ y)\cdot z = x\cdot z + y\cdot z$\\
            $A5$ & $(x\cdot y)\cdot z = x\cdot(y\cdot z)$\\
            $P1$ & $x\parallel y = y \parallel x$\\
            $P2$ & $(x\parallel y)\parallel z = x\parallel (y\parallel z)$\\
            $P3$ & $e_1\parallel (e_2\cdot y) = (e_1\parallel e_2)\cdot y$\\
            $P4$ & $(e_1\cdot x)\parallel e_2 = (e_1\parallel e_2)\cdot x$\\
            $P5$ & $(e_1\cdot x)\parallel (e_2\cdot y) = (e_1\parallel e_2)\cdot (x\parallel y)$\\
            $P6$ & $(x+ y)\parallel z = (x\parallel z)+ (y\parallel z)$\\
            $P7$ & $x\parallel (y+ z) = (x\parallel y)+ (x\parallel z)$\\
        \end{tabular}
        \caption{Axioms of PA1}
        \label{AxiomsForPA1}
    \end{table}
\end{center}

We can get the following soundness and completeness theorems of PA1 modulo pomset, step and hp-bisimulations, for details, please refer to APTC \cite{APTC}.

\begin{theorem}[Soundness of PA1 modulo truly concurrent bisimulation equivalences]\label{SPA}
The axiomatization of PA1 is sound modulo truly concurrent bisimulation equivalences $\sim_{p}$, $\sim_{s}$, and $\sim_{hp}$. That is,

\begin{enumerate}
  \item let $x$ and $y$ be PA1 terms. If PA1 $\vdash x=y$, then $x\sim_{p} y$;
  \item let $x$ and $y$ be PA1 terms. If PA1 $\vdash x=y$, then $x\sim_{s} y$;
  \item let $x$ and $y$ be PA1 terms. If PA1 $\vdash x=y$, then $x\sim_{hp} y$.
\end{enumerate}

\end{theorem}

\begin{theorem}[Completeness of PA1 modulo truly concurrent bisimulation equivalences]\label{CPA}
The axiomatization of PA1 is complete modulo truly concurrent bisimulation equivalences $\sim_{p}$, $\sim_{s}$, and $\sim_{hp}$. That is,

\begin{enumerate}
  \item let $p$ and $q$ be closed PA1 terms, if $p\sim_{p} q$ then $p=q$;
  \item let $p$ and $q$ be closed PA1 terms, if $p\sim_{s} q$ then $p=q$;
  \item let $p$ and $q$ be closed PA1 terms, if $p\sim_{hp} q$ then $p=q$.
\end{enumerate}

\end{theorem}

\section{On Axiomatization of Parallel Composition for Hereditary History-Preserving Bisimulation}{\label{oahhp}}

The axioms in Table \ref{AxiomsForPA1} is not sound modulo hhp-bisimulation. And further, since hhp-bisimulation is downward-closed hp-bisimulation and can be downward to atomic events, 
and implies bisimulation. According to the axiomatization work of the parallel composition modulo bisimulation \cite{ILM} \cite{HM} \cite{PC}, the following negative conclusions still 
hold for the parallel composition modulo hhp-bisimulation.

\begin{proposition}
PA1 does not have a finite sound and complete axiomatization modulo hhp-bisimulation.
\end{proposition}

\section{Axiomatization of Parallel Composition for Hereditary History-Preserving Bisimulation}{\label{ahhp}}

Since the finite sound and complete axiomatization of the parallel composition for hhp-bisimulation does not exist, we introduce two auxiliary operators: the left parallel composition
$\leftmerge$ and the communication merge $\mid$. The PA in this section is called PA2. We give the transition rules of the left parallel composition and communication merge in Table \ref{TRForPA2}, 
it is suitable for all truly concurrent behavioral equivalence, including pomset bisimulation, step bisimulation, hp-bisimulation and hhp-bisimulation.

\begin{center}
    \begin{table}
        $$\frac{}{e\xrightarrow{e}\surd}$$
        $$\frac{x\xrightarrow{e}\surd}{x+ y\xrightarrow{e}\surd} \quad\frac{x\xrightarrow{e}x'}{x+ y\xrightarrow{e}x'} \quad\frac{y\xrightarrow{e}\surd}{x+ y\xrightarrow{e}\surd}
        \quad\frac{y\xrightarrow{e}y'}{x+ y\xrightarrow{e}y'}$$
        $$\frac{x\xrightarrow{e}\surd}{x\cdot y\xrightarrow{e} y} \quad\frac{x\xrightarrow{e}x'}{x\cdot y\xrightarrow{e}x'\cdot y}$$
        $$\frac{x\xrightarrow{e_1}\surd\quad y\xrightarrow{e_2}\surd \quad(e_1\leq e_2)}{x\leftmerge y\xrightarrow{\{e_1,e_2\}}\surd} \quad\frac{x\xrightarrow{e_1}x'\quad y\xrightarrow{e_2}\surd \quad(e_1\leq e_2)}{x\leftmerge y\xrightarrow{\{e_1,e_2\}}x'}$$
        $$\frac{x\xrightarrow{e_1}\surd\quad y\xrightarrow{e_2}y' \quad(e_1\leq e_2)}{x\leftmerge y\xrightarrow{\{e_1,e_2\}}y'} \quad\frac{x\xrightarrow{e_1}x'\quad y\xrightarrow{e_2}y' \quad(e_1\leq e_2)}{x\leftmerge y\xrightarrow{\{e_1,e_2\}}x'\parallel y'}$$
        $$\frac{x\xrightarrow{e_1}\surd\quad y\xrightarrow{e_2}\surd}{x\mid y\xrightarrow{\gamma(e_1,e_2)}\surd} \quad\frac{x\xrightarrow{e_1}x'\quad y\xrightarrow{e_2}\surd}{x\mid y\xrightarrow{\gamma(e_1,e_2)}x'}$$
        $$\frac{x\xrightarrow{e_1}\surd\quad y\xrightarrow{e_2}y'}{x\mid y\xrightarrow{\gamma(e_1,e_2)}y'} \quad\frac{x\xrightarrow{e_1}x'\quad y\xrightarrow{e_2}y'}{x\mid y\xrightarrow{\gamma(e_1,e_2)}x'\parallel y'}$$
        \caption{Transition rules of PA2}
        \label{TRForPA2}
    \end{table}
\end{center}

We can design the axioms of PA2 as Table \ref{AxiomsForPA2} shows.

\begin{center}
    \begin{table}
        \begin{tabular}{@{}ll@{}}
            \hline No. &Axiom\\
            $A1$ & $x+ y = y+ x$\\
            $A2$ & $(x+ y)+ z = x+ (y+ z)$\\
            $A3$ & $x+ x = x$\\
            $A4$ & $(x+ y)\cdot z = x\cdot z + y\cdot z$\\
            $A5$ & $(x\cdot y)\cdot z = x\cdot(y\cdot z)$\\
            $P1$ & $x\parallel y = x\leftmerge y + y\leftmerge x + x\mid y$\\
            $L2$ & $(e_1\leq e_2)\quad e_1\leftmerge (e_2\cdot y) = (e_1\leftmerge e_2)\cdot y$\\
            $L3$ & $(e_1\leq e_2)\quad (e_1\cdot x)\leftmerge e_2 = (e_1\leftmerge e_2)\cdot x$\\
            $L4$ & $(e_1\leq e_2)\quad (e_1\cdot x)\leftmerge (e_2\cdot y) = (e_1\leftmerge e_2)\cdot (x\parallel y)$\\
            $L5$ & $(x+ y)\leftmerge z = (x\leftmerge z)+ (y\leftmerge z)$\\
            $C6$ & $e_1\mid e_2 = \gamma(e_1,e_2)$\\
            $C7$ & $e_1\mid (e_2\cdot y) = \gamma(e_1,e_2)\cdot y$\\
            $C8$ & $(e_1\cdot x)\mid e_2 = \gamma(e_1,e_2)\cdot x$\\
            $C9$ & $(e_1\cdot x)\mid (e_2\cdot y) = \gamma(e_1,e_2)\cdot (x\between y)$\\
            $C10$ & $(x+ y)\mid z = (x\mid z) + (y\mid z)$\\
            $C11$ & $x\mid (y+ z) = (x\mid y)+ (x\mid z)$\\
        \end{tabular}
        \caption{Axioms of PA2}
        \label{AxiomsForPA2}
    \end{table}
\end{center}

Then we can get the following soundness and completeness theorems of PA2 modulo pomset, step, hp-, and hhp-bisimulations, for details, please refer to APTC \cite{APTC}.

\begin{theorem}[Soundness of PA2 modulo truly concurrent bisimulation equivalences]\label{SPA}
The axiomatization of PA2 is sound modulo truly concurrent bisimulation equivalences $\sim_{p}$, $\sim_{s}$, $\sim_{hp}$ and $\sim_{hhp}$. That is,

\begin{enumerate}
  \item let $x$ and $y$ be PA2 terms. If PA2 $\vdash x=y$, then $x\sim_{p} y$;
  \item let $x$ and $y$ be PA2 terms. If PA2 $\vdash x=y$, then $x\sim_{s} y$;
  \item let $x$ and $y$ be PA2 terms. If PA2 $\vdash x=y$, then $x\sim_{hp} y$;
  \item let $x$ and $y$ be PA2 terms. If PA2 $\vdash x=y$, then $x\sim_{hhp} y$.
\end{enumerate}

\end{theorem}

\begin{theorem}[Completeness of PA2 modulo truly concurrent bisimulation equivalences]\label{CPA}
The axiomatization of PA2 is complete modulo truly concurrent bisimulation equivalences $\sim_{p}$, $\sim_{s}$, $\sim_{hp}$, and $\sim_{hhp}$. That is,

\begin{enumerate}
  \item let $p$ and $q$ be closed PA2 terms, if $p\sim_{p} q$ then $p=q$;
  \item let $p$ and $q$ be closed PA2 terms, if $p\sim_{s} q$ then $p=q$;
  \item let $p$ and $q$ be closed PA2 terms, if $p\sim_{hp} q$ then $p=q$;
  \item let $p$ and $q$ be closed PA2 terms, if $p\sim_{hhp} q$ then $p=q$.
\end{enumerate}

\end{theorem}

\section{Conclusions}{\label{con}}

We recall the parallel composition for true concurrency as sidelights of APTC, and prove that:
\begin{enumerate}
  \item There is a finite sound and complete axiomatization of the parallel composition modulo pomset, step and hp-bisimulations, without any auxiliary operators.
  \item There does not exist a finite sound and complete axiomatization of the parallel composition modulo hhp-bisimulation, without any auxiliary operator.
  \item There is a finite sound and complete axiomatization of the parallel composition modulo pomset, step, hp-. and hhp-bisimulations, with the auxiliary left parallel composition
  and communication merge.
\end{enumerate}

\newpage

%

\label{lastpage}

\end{document}